\documentclass[prb,twocolumn,showpacs,preprintnumbers,amsmath,amssymb]{revtex4}
\usepackage{graphicx}% Include figure files with \includegraphics
\usepackage{epsfig}% Include figure files with \psfig
\begin{document}
\title{Stability of condensate in superconductors}
\author{P. Lipavsk\'y$^{1,2}$, B. {\v S}op{\'\i}k$^{1,2}$, M. M{\"a}nnel$^3$ and K. Morawetz$^{4,5}$}
\address{$^1$Faculty of Mathematics and Physics,
Charles University, Ke Karlovu 3, 12116 Prague 2, Czech Republic}
\address{$^2$Institute of Physics, Academy of Sciences,
Cukrovarnick\'a 10, 16253 Prague 6, Czech Republic}
\address{$^3$Institute of Physics, Chemnitz University of Technology, 09107 Chemnitz, Germany}
\address{$^4$International Center of Condensed Matter Physics, University of Brasilia, 70904-970, Brasilia-DF, Brazil}
\address{$^5$M{\"u}nster University of Applied Science, Stegerwaldstrasse 39, 48565 Steinfurt, Germany}

\begin{abstract}
According to the BCS theory the superconducting condensate develops 
in a single quantum mode and no Cooper pairs out of the condensate
are assumed. Here we discuss a mechanism by which 
the successful mode inhibits condensation in neighboring modes and
suppresses a creation of noncondensed Cooper pairs. It is shown that
condensed and noncondensed Cooper pairs are separated by an energy 
gap which is smaller than the superconducting gap but large enough 
to prevent nucleation in all other modes and to eliminate effects 
of noncondensed Cooper pairs on properties of superconductors.
Our result thus justifies basic assumptions of the BCS theory and
confirms that the BCS condensate is stable with respect to 
two-particle excitations.
\end{abstract}

\maketitle

\section{Introduction}
Seven decades ago London put forward the idea to explain 
superconductivity by a rigid quantum wave function covering 
all superconducting electrons. His bold vision is included 
in all recent theories would the role of the wave function 
be fulfilled by the Ginzburg-Landau (GL) complex order parameter
or the Bardeen-Cooper-Schrieffer (BCS) gap function \cite{BCS57,Gor59}. 
The superconducting condensate is not called rigid but it is
assumed to be sufficiently stable to form an effective vacuum 
of a new state, the state of broken symmetry \cite{Goldstone61}. 
This stability is always used but its origin is not yet clear. 

In superconductors the macroscopic wave function is a 
phenomenological theoretical tool with a complicated 
underlying microscopic picture. In contrast, the macroscopic 
wave function of dilute superfluid gases is identical to the 
intuitively clear Schr{\"o}dinger wave function of the lowest 
energy single-particle state macroscopically occupied due to 
the Bose-Einstein condensation.\cite{SG98} On superfluids we 
can outline the problem addressed here for superconductors.

Landau has shown that the supercurrent is halted if external 
perturbations excite bosons from the condensate to neighboring 
low-laying states.\cite{PS01} According to this Landau 
criterion a supercurrent in the ideal (noninteracting) Bose gas 
is unstable. Indeed, in the ideal Bose gas the energy of a 
neighbor state differs by the kinetic energy which is quadratic 
in momentum of this excitation. For any slow perturbation one 
then finds states of slower velocity vulnerable to excitation 
by Cherenkov-like mechanism. In real Bose systems the situation 
is different. The interaction between bosons causes a 
reconstruction of the energy spectrum from the 
quadratic to the linear form of acoustic type. Perturbations 
slower than the corresponding sound velocity then cannot 
excite particles so that the supercurrent is stable and the 
condensate wave function reveals the London rigidity.

Excitations of the superconductors are principally different. 
According to the BCS theory well supported by experimental 
experience, see e.g. Ref.~\onlinecite{PWVHW82}, the 
superconductivity is controlled by fermionic quasiparticles 
resulting from broken Cooper pairs. We don't discuss here 
this familiar mechanism. Our central question
is: {\em Why Cooper pairs of nonzero momenta are not excited
from the BCS condensate?}

Here we show that the multiple scattering corrections to the 
T-matrix \cite{Lipavsky08} lead to gaps in the single-particle 
and two-particle energy spectra. The single-particle gap is
the familiar BCS gap which is known to guarantee stability with 
respect to the excitation of fermionic quasiparticles. We
focus on the two-particle gap and show that it guarantees
stability with respect to nucleation of superconducting 
condensate in two or more momentum states and with respect
to excitation of noncondensed Cooper pairs. These two gaps
thus imply the stable condensate.

The paper is organized as follows. In Sec.~\ref{TA} we introduce
the T-matrix approach first on a general level, Sec.~\ref{GE},
and then its simplified form for the separable interaction of
the BCS type, Sec.~\ref{SI}. Then we evaluate the T-matrix near
the critical temperature for the condensate in Sec.~\ref{CM},
and for the noncondensed Cooper pairs in Sec.~\ref{NCP}. In
Sec.~\ref{SC} we discuss the stability of the condensate with
respect to nucleation of a second condensate, Sec.~\ref{EPC},
and with respect to excitation of noncondensed Cooper pairs,
Sec.~\ref{ECP}. Comparing the present approximation with 
the Kadanoff-Martin theory in Sec.~\ref{MSC} we point out the 
role of multiple scattering correction in formation of the gap
in the two-particle energy spectrum. Section \ref{Sum} is a
summary.

\section{T-matrix approach}\label{TA}
The noncondensed Cooper pairs are not covered by the BCS or 
Eliashberg theory. These approaches treat Cooper pairs within
the mean field which is nonzero only for states with macroscopic
bosonic occupation -- the states with the condensate.
In our study we employ the T-matrix in which the pairing is 
independent of the condensation. We use the Galitskii-Feynman 
approximation with multiple scattering corrections\cite{Lipavsky08}
in modification of Ref.~\onlinecite{SLMM10}. 

\subsection{General equations}\label{GE}
Let us introduce the theory.
The full propagator is given by the Dyson equation 
\begin{equation}
G_{\uparrow}=G^0+G^0\Sigma_{\uparrow}G_{\uparrow},
\label{G}
\end{equation}
where the selfenergy 
\begin{equation}
\Sigma_{\uparrow}={k_{\rm B}T\over\Omega}\sum_{Q}
\sigma_{Q\uparrow}
\label{S}
\end{equation}
is a sum over four-momentum $Q\equiv (\omega,{\bf Q})$ of 
interacting pairs, with Matsubara's frequencies 
$\omega$ and discrete wave vectors $\bf Q$ corresponding 
to the sample volume $\Omega$. In the case of condensation 
mode $Q$ is a four-momentum of a Cooper pair.
$\sigma_{Q\uparrow}$ we call a $Q$-part.

To avoid double-counts, the internal lines of the $Q$-part 
of the selfenergy should not include processes related 
to the $Q$-mode itself. To this end we introduce the 
$Q$-reduced propagator
\begin{equation}
G_{{\not Q}\downarrow}=G^0+G^0\Sigma_{{\not Q}\downarrow}G_{{\not Q}\downarrow},
\label{Gq}
\end{equation}
which is dressed by all but the $Q$-part of the selfenergy
\begin{equation}
\Sigma_{{\not Q}\downarrow}={k_{\rm B}T\over\Omega}\sum_{Q'\ne Q}
\sigma_{Q'\downarrow}. 
\label{Sq}
\end{equation}

The $Q$-part of the $\uparrow$ selfenergy is obtained by 
closing the loop of $\downarrow$ line of the T-matrix by 
the reduced propagator
\begin{equation}
\sigma_{Q\uparrow}(k)={\cal T}_{\uparrow\downarrow}
(k,Q\!-\!k;k,Q\!-\!k)G_{{\not Q}\downarrow}(Q\!-\!k).
\label{Sq}
\end{equation}

The T-matrix is constructed from 
the $Q$-reduced propagators in the $\downarrow$ line and full 
propagators in the $\uparrow$ line
\begin{eqnarray}
&&{\cal T}_{\uparrow\downarrow}(k,Q\!-\!k;p,Q\!-\!p)=
D(k,Q\!-\!k;p,Q\!-\!p)
\nonumber\\
&&-{k_{\rm B}T\over\Omega}{\sum_{k'}} 
D(k,Q\!-\!k;k',Q\!-\!k')
G_{\uparrow}(k')
G_{{\not Q}\downarrow}(Q\!-\!k')
\nonumber\\
&&~~~~~~~~~~~~~~~~\times
{\cal T}_{\uparrow\downarrow}(k',Q\!-\!k';p,Q\!-\!p).
\label{Tq}
\end{eqnarray}
Except for the reduced propagator, this is the standard ladder
approximation. 
Here $D$ can be either a general phonon propagator with vortices 
included or an effective interaction potential of the BCS type.
The set of equations is complete. 

The reduced propagator eliminates nonphysical repeated 
collisions in the spirit of multiple scattering expansion.
If repeated collisions are not eliminated, which is achieved 
using approximation $G_{{\not Q}\downarrow}\approx G_{\downarrow}$, 
one recovers the original Galitskii-Feynman approximation. 
Importance of the multiple scattering corrections can be
seen from properties of the original Galitskii-Feynman 
approximation in the superconducting state. The T-matrix 
becomes singular which signals the onset of pairing. The 
single-particle propagator, however, does not have the gap
in the energy spectrum.\cite{Wild60,KM61} With the multiple
scattering corrections the gap develops.\cite{Lipavsky08,SLMM10}

The half-selfconsistent theory \cite{KM61} of Kadanoff and 
Martin (KM) is recovered if we approximate the $Q$-reduced 
propagator by the bare one, $G_{{\not Q}\downarrow}\approx G^0$. 
The KM theory yields the correct BCS gap but as 
noticed by Chen {\em et al} \cite{CSTL05} and confirmed below, 
the KM approximation results in the ideal Bose gas of Cooper 
pairs. According to the Landau criterion the KM theory does 
not explain stability of the condensate with respect to 
excitation of noncondensed Cooper pairs. The reduced 
selfconsistency of the multiple scattering 
approach is thus essential for the excitation spectrum.

\subsection{Separable interaction}\label{SI}
For a discussion in this paper we employ the simple 
BCS interaction
\begin{eqnarray}
&D(k,Q\!-\!k;p,Q\!-\!p)
=-V
\theta\!\left(\omega_{\rm D}\!-\!\left|\epsilon(k)\right|\right)
\nonumber\\
&\times\theta\!\left(\omega_{\rm D}\!-\!\left|\epsilon(Q\!-\!k)\right|\right)
\theta\!\left(\omega_{\rm D}\!-\!\left|\epsilon(p)\right|\right)
\theta\!\left(\omega_{\rm D}\!-\!\left|\epsilon(Q\!-\!p)\right|\right)\!.
\label{BCS-V}
\end{eqnarray}
From Eq.~\eqref{Tq} one can see that this separable potential 
implies the separable T-matrix
\begin{eqnarray}
&{\cal T}_{\uparrow\downarrow}(k,Q\!-\!k;p,Q\!-\!p)
=-{\cal T}_{Q}
\theta\!\left(\omega_{\rm D}\!-\!\left|\epsilon(k)\right|\right)
\nonumber\\
&\times\theta\!\left(\omega_{\rm D}\!-\!\left|\epsilon(Q\!-\!k)\right|\right)
\theta\!\left(\omega_{\rm D}\!-\!\left|\epsilon(p)\right|\right)
\theta\!\left(\omega_{\rm D}\!-\!\left|\epsilon(Q\!-\!p)\right|\right)
\label{BCS-T}
\end{eqnarray}
and Eq.~\eqref{Tq} simplifies to a scalar equation
\begin{equation}
{1\over{\cal T}_{Q}}={1\over V}+{k_{\rm B}T\over\Omega}{\sum_{k}} 
G_{\uparrow}(k)
G_{{\not Q}\downarrow}(Q\!-\!k).
\label{Ti}
\end{equation}
The sum over $k$ is restricted by cutoffs of the BCS model.
The $Q$-part of the selfenergy simplifies to
\begin{equation}
\sigma_{Q\uparrow}(k)=-{\cal T}_{Q}
G_{{\not Q}\downarrow}(Q\!-\!k) 
\theta\left(\omega_{\rm D}\!-\!\left|\epsilon(k)\right|\right).
\label{BCS-Sq}
\end{equation}
Equations \eqref{G}-\eqref{Gq} and \eqref{Ti}-\eqref{BCS-Sq}
form a closed set.

\subsection{Condensation mode}\label{CM}
The condensation of Cooper pairs happens in a single \mbox{mode}. 
Below we prove this assumption. From now on we reserve the index 
$Q$ for the condensation mode, while the other modes will be 
denoted by $Q'$. The Matsubara frequency in $Q$ is zero, 
$Q=(0,{\bf Q})$.

The T-matrix of the $Q$-mode diverges reaching values 
proportional to the volume $\Omega$, see Ref.~\onlinecite{Lipavsky08}. 
To make a link with the standard notation of the BCS 
theory we express this singular element as
\begin{equation}
{\cal T}_Q={\Omega\over k_{\rm B}T}\bar\Delta\Delta,
\label{Deldef}
\end{equation}
and split the selfenergy into the singular contribution of the
$Q$-mode and the regular reminder 
\begin{eqnarray}
\Sigma_{\uparrow}(k)&=&-\bar\Delta G_{{\not Q}\downarrow}(-k)\Delta+
{k_{\rm B}T\over\Omega}\sum_{Q'\not= Q}\sigma_{Q'\uparrow}(k)
\nonumber\\
&=&-\bar\Delta G_{{\not Q}\downarrow}(-k)\Delta+\Sigma_{{\not Q}\uparrow}(k).
\label{Ss}
\end{eqnarray}
According to Eqs.~\eqref{G} and~\eqref{Gq}, the $Q$-reduced 
propagator relates to the full propagator as 
\begin{equation}
G_{{\not Q}\downarrow}=G_{\downarrow}+G_{\downarrow}\bar\Delta 
G_{{\not Q}\uparrow}\Delta G_{{\not Q}\downarrow}.
\label{Gn0}
\end{equation}
If one neglects renormalizations keeping only the gap, 
$\Sigma_{\uparrow}\approx-\bar\Delta G_{{\not Q}\downarrow}\Delta$,
i.e., $\Sigma_{{\not Q}\uparrow}\approx 0$ or $G_{{\not Q}\downarrow}
\approx G^0$, this equation becomes 
identical to the Nambu-Gor'kov equation with the BCS gap $\Delta$. 

The T-matrix of the $Q$-mode 
\begin{equation}
{1\over{\cal T}_{Q}}={1\over V}+{k_{\rm B}T\over\Omega}{\sum_{k}} 
G_{\uparrow}(k)
G_{{\not Q}\downarrow}(Q\!-\!k)
\label{T0} 
\end{equation}
determines the gap. In the thermodynamical limit $\Omega\to\infty$, 
the T-matrix of the condensation mode diverges, i.e., 
$1/{\cal T}_{Q}\to 0$. Equation~\eqref{T0} then simplifies to the 
BCS-like gap equation 
\begin{equation}
0={1\over V}+{k_{\rm B}T\over\Omega}{\sum_{k}} 
G_{\uparrow}(k)
G_{{\not Q}\downarrow}(Q\!-\!k).
\label{T0lim}
\end{equation}

Gor'kov have analyzed equation \eqref{T0lim} close to the critical 
temperature $T_c$, where the gap is small. Keeping terms to the 
quadratic order in $\Delta$ he has shown that it leads to
\begin{equation}
{{\bf Q}^2\over 2m^*}+\alpha+\beta|\Delta|^2=0,
\label{GL1}
\end{equation}
where $m^*$ is a Cooperon mass, $\beta=3/
(2E_{\rm F})$ and $\alpha=-6\pi^2k_{\rm B}^2T_c(T_c-T)/
(7\zeta_{[3]}E_{\rm F})  $ are 
the GL parameters with $n$ being the electron density, $E_{\rm F}$ 
the Fermi energy, and $\zeta_{[3]}=1.202$ the Riemann zeta function. 
%These formulas have been derived for a parabolic electron bands, 
%however, one can express the Fermi energy via the density of 
%states, $E_{\rm F}={4\over 3}nN_0$, and use realistic $N_0$.
We restrict our attention to the vicinity of the critical 
temperature. The limiting form \eqref{GL1} will be thus sufficient 
for our discussion. 

\subsection{Non-condensed Cooper pairs}\label{NCP}
We expect that none of terms for $Q'\not= Q$ diverges with volume. 
This expectation is confirmed below. The $Q'$-reduced propagator 
then approaches the full one in the thermodynamical limit
$\Omega\to\infty$,
\begin{equation}
G_{{\not Q'}\uparrow}=G_{\uparrow}.
\label{Gqs}
\end{equation}
and the T-matrix of a $Q'\not=Q$-mode 
\begin{equation}
{1\over{\cal T}_{Q'}}={1\over V}+{k_{\rm B}T\over\Omega}{\sum_{k}} 
G_{\uparrow}(k)
G_{\downarrow}(Q'\!-\!k)
\label{Tqn0}
\end{equation}
thus satisfies equation distinct from Eq.~\eqref{T0}. 
For the condensation mode \eqref{T0} the 
gap enters only one of propagators while for noncondensation 
modes \eqref{Tqn0} both propagators depend on the gap. This difference
results in a suppressed excitation of noncondensed Copper pairs.

The inverse T-matrix of noncondensation mode $Q'=(0,{\bf Q'})$ 
results from expansion of Eq.~\eqref{Tqn0} to the quadratic 
order in $\Delta$ as
\begin{equation}
{C\over{\cal T}_{Q'}}=
{{\bf Q'}^2\over 2m^*}-|\alpha|+2\beta|\Delta|^2,
\label{GL2}
\end{equation}
where $C=8\pi^2k_{\rm B}^2T_c^2/(7\zeta_{[3]}n)$. The factor of two
in front of $\beta$ follows from the fact that for noncondensed 
pairs both propagators depend on the gap. 

\section{Stability of the condensate}\label{SC}
Now we are ready to solve the central problem of this paper. 
First we show that once the condensate is formed in the 
$Q$-mode, a parallel condensation in another $Q'$-mode is 
excluded. Second we show that the critical velocity for breaking the 
condensed Cooper into two quasiparticles is lower than the
critical velocity of excitation of Cooper pairs out of condensate.

\subsection{Excluded parallel condensation}\label{EPC}
The inverse T-matrix \eqref{GL2} of the non-condensation $Q'$-mode 
cannot reach zero turning the $Q'$-mode into a parallel condensation
mode. To show this
we first use the GL equation \eqref{GL1}, to express the T-matrix 
of noncondensation mode as
\begin{equation}
{C\over{\cal T}_{Q'}}=
{{\bf Q'}^2\over 2m^*}+|\alpha|-{{\bf Q}^2\over m^*}.
\label{GL3}
\end{equation}

Values of the pair momentum $\bf Q$ are limited by the critical
current, ${\bf Q}^2<{\bf Q}_c^2$. The current is proportional to 
the square of the gap times the momentum, ${\bf j}\propto {\bf Q}
|\Delta|^2$. Using Eq. \eqref{GL1} one finds ${\bf j}\propto 
{\bf Q}\left(|\alpha|-{\bf Q}^2/2m^*\right)$. The critical current 
is the maximum one, $\partial{\bf j}/\partial
{\bf Q}|_{{\bf Q}_c}=0$, which is achieved for 
${\bf Q}_c^2=2m^*|\alpha|/3$, see Tinkham\cite{Tinkham}. 
Accordingly,
\begin{equation}
{C\over{\cal T}_{Q'}}>
{{\bf Q'}^2\over 2m^*}+|\alpha|-{{\bf Q}_c^2\over m^*}=
{{\bf Q'}^2\over 2m^*}+{|\alpha|\over 3}.
\label{GL4}
\end{equation}

Inequality \eqref{GL4} implies that the mode of $Q'\not= Q$ 
cannot become singular once the condensation develops in the 
mode $Q$. Therefore, a parallel condensation in two competitive 
modes is excluded. Briefly, there is a single condensate, as it 
is tacitly assumed in the BCS theory.

\subsection{Excitation of Cooper pairs from the condensate}\label{ECP}
Now we discuss a possibility to excite a Cooper pair out of
condensate by an object moving with velocity $\bf v$ in the 
static condensate. Going into the floating coordinate system,
this criterion is used to check stability of the condensate 
flowing with velocity $-\bf v$ around a static obstacle. 

The right hand side of Eq. \eqref{GL2} represents an energy of
a noncondensed Cooper pair of momentum $Q'$. In the frame floating 
with the condensate, ${\bf Q=0}$, a Cooper pair can be excited 
from the condensate into a noncondensed state with the minimal 
energy cost $|\alpha|$. Let us estimate under which conditions
Cooper pairs can be excited by an external perturbation.

According to the Landau criterion \cite{PS01} the external 
perturbation moving with velocity ${\bf v}$ can 
excite the Cooper pair of momentum $\bf Q'$ if the Cherenkov
condition 
\begin{equation}
{\bf vQ'}={{\bf Q'}^2\over 2m^*}+|\alpha|
\label{Cherenkov}
\end{equation}
is satisfied. This equation is solved by real $\bf Q'$ for
\begin{equation}
|{\bf v}|>\sqrt{2|\alpha|\over m^*}.
\label{GL5}
\end{equation}
This velocity is higher than the critical velocity of pair 
breaking $v_c=\Delta/k_{\rm F}$, where $k_{\rm F}$ is the 
Fermi momentum. Indeed, from Eq. \eqref{GL1} follows 
$\Delta=\sqrt{|\alpha|/\beta}=\sqrt{|\alpha| k_{\rm F}^2/(3m)}$, 
where $m=m^*/2$ is the electronic mass, therefore
\begin{equation}
|{\bf v}|>\sqrt{3}v_c.
\label{GL6}
\end{equation}
Briefly, it is easier to break a Cooper pair into two
quasiparticles than to excite it from the condensate 
into a noncondensed Cooper pair.

%One comment is necessary. It is known from Bose-Einstein
%condensation (BEC) that the excitation spectrum of 
%noncondensed bosons has acoustic form being linear in 
%the momentum. In fact, the quadratic spectrum with a gap 
%easily results as an artifact of non-conserving 
%approximations \cite{SG98}. We believe that spectrum 
%obtained here is not an artifact, because the present 
%analysis applies to the BCS side of the BEC-BCS crossover 
%while the picture of interacting bosons applies to the 
%BEC side. 

\subsection{Role of the multiple scattering corrections}\label{MSC}
As pointed out above deriving Eqs.~\eqref{T0} and \eqref{Tqn0}, 
within the multiple scattering approach the propagators
inside the T-matrix depend on the evaluated scattering
process. As the condensation mode becomes singular, the
two-particle propagations in the condensed and noncondensed 
modes become particularly different. 
Let us show that this difference is essential using the
Kadanoff-Martin approximation which results from the present
approximation using $G_{{\not Q}\downarrow}\approx G^0$,
therefore it uses the same two-particle
propagation $G_{\uparrow}(k)G^0(Q-k)$ for all modes.

Now we confirm that the KM approximation results 
in the ideal gas of Cooper pairs \cite{CSTL05}. Since one 
of propagators is bare for all modes, the KM counterpart 
of equation~\eqref{GL2} reads ${C\over{\cal T}_{Q'}}=
{{\bf Q'}^2\over 2m^*}+|\alpha|+\beta|\Delta|^2$ so that 
equation \eqref{GL3} modifies to ${C\over{\cal T}_{Q'}}=
{{\bf Q'}^2\over 2m^*}-{{\bf Q}^2\over 2m^*}$. When the
condensate moves, ${\bf Q}\not= 0$, it is energetically 
favorable to start condensation in standing mode 
${\bf Q}'=0$ which stops the supercurrent. A similar 
problem appears for noncondensed Cooper pairs. In the frame
moving with the condensate one finds the free-particle-like 
energy of noncondensed Cooper pairs ${{\bf Q'}^2/ 2m^*}$. 
Therefore, according to the Landau criterion in the KM 
approximation the condensate is not stable.

\section{Summary}\label{Sum}
We have discussed stability of supercurrents 
with respect to condensation in competitive modes and 
excitations of noncondensed Cooper pairs. It was shown 
within the Galitskii-Feynman approximation that multiple 
scattering corrections yield the familiar BCS gap in the 
single-particle energy spectrum and also a smaller 
gap in the two-particle energy spectrum separating 
the noncondensed Cooper pairs from the condensate. This 
two-particle gap prevents parallel condensation of Cooper 
pairs in two or more modes. Moreover, due to the two-particle 
gap the critical velocity to excite noncondensed Cooper pairs 
is higher than the critical velocity of the pair breaking, 
therefore the noncondensed Cooper pairs do not affect 
stability of supercurrents. The present result justifies
basic assumptions of the BCS theory in which the condensate
is expected in a single mode and Cooper pairs out of the
condensate are ignored.

\acknowledgements
This work was supported by research plans MSM 0021620834, 
grants GAUK 135909, GA\v{C}R 204/10/0687 and 202/08/0326, 
GAAV 100100712, DAAD PPP, and by DGF-CNPq project 
444BRA-113/57/0-1.

\bibliography{bose,delay2,delay3,gdr,genn,chaos,kmsr,kmsr1,kmsr2,kmsr3,kmsr4,kmsr5,kmsr6,kmsr7,micha,refer,sem1,sem2,sem3,short,spin,spin1,solid,deform,paradox,tdgl}

\end{document}